\documentclass[10pt]{article}
\usepackage{psfig}
\newcommand{\figcapskip}{-0.5\baselineskip}
\begin{document}

\title{Thermalization at lowest energies? \\
A view from a transport model }

\author{C Hartnack\footnote{invited speaker, email hartnack@subatech.in2p3.fr}
$^1$, H Oeschler$^2$
and J Aichelin$^1$ \\[1ex]
$^1$SUBATECH,
4, rue A. Kastler, 
44307 Nantes, France \\
$^2$ Inst. f. Kernphysik, TU Darmstadt
64289 Darmstadt, Germany}

\date{}
\maketitle

\begin{abstract}
Using the Isospin Quantum Molecular Dynamics (IQMD) model we
analyzed the production of pions and kaons in the energy range of
1-2 $A$GeV in order to study the question why thermal models could
achieve a successful description. For this purpose we study the
variation of pion and kaon yields using different elementary cross
sections. We show that several ratios appear to be rather robust
versus their variations.
\end{abstract}

\vspace*{-3mm}
\section{Introduction}

In the last decades heavy ion experiments have studied various
combinations of heavy ion collisions. Besides the understanding of
dynamical observables the production of particles have been of a
special focus. Recent experiments at the SIS accelerator at GSI
have us provided with high-quality data of the production of pions
\cite{reisdorf} and of kaons \cite{Devismes,FoersterPRC}. The
yields of particle ratios have been shown of large interest
especially for the application of thermal models \cite{cleymans}.
Several ratios of measured yields could be interpreted by the use
of law-of-mass relations \cite{ho_s2000,law-of-mass}. That idea
was later on supported by a detailed analysis from transport
models~\cite{kminus,bigpr}.

\begin{figure}[hbt]
\centerline{\psfig{file=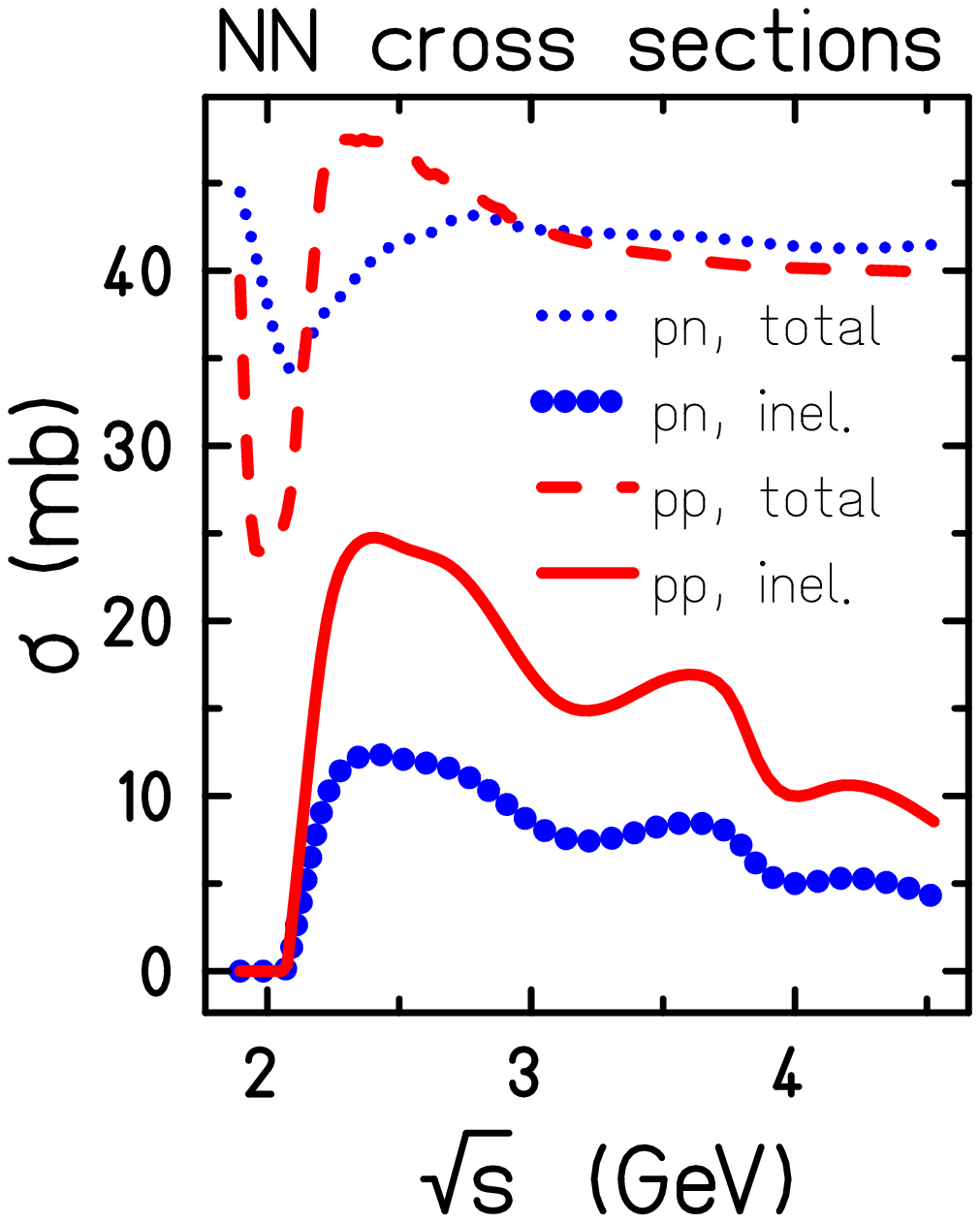,width=0.35\textwidth,angle=-0}
\psfig{file=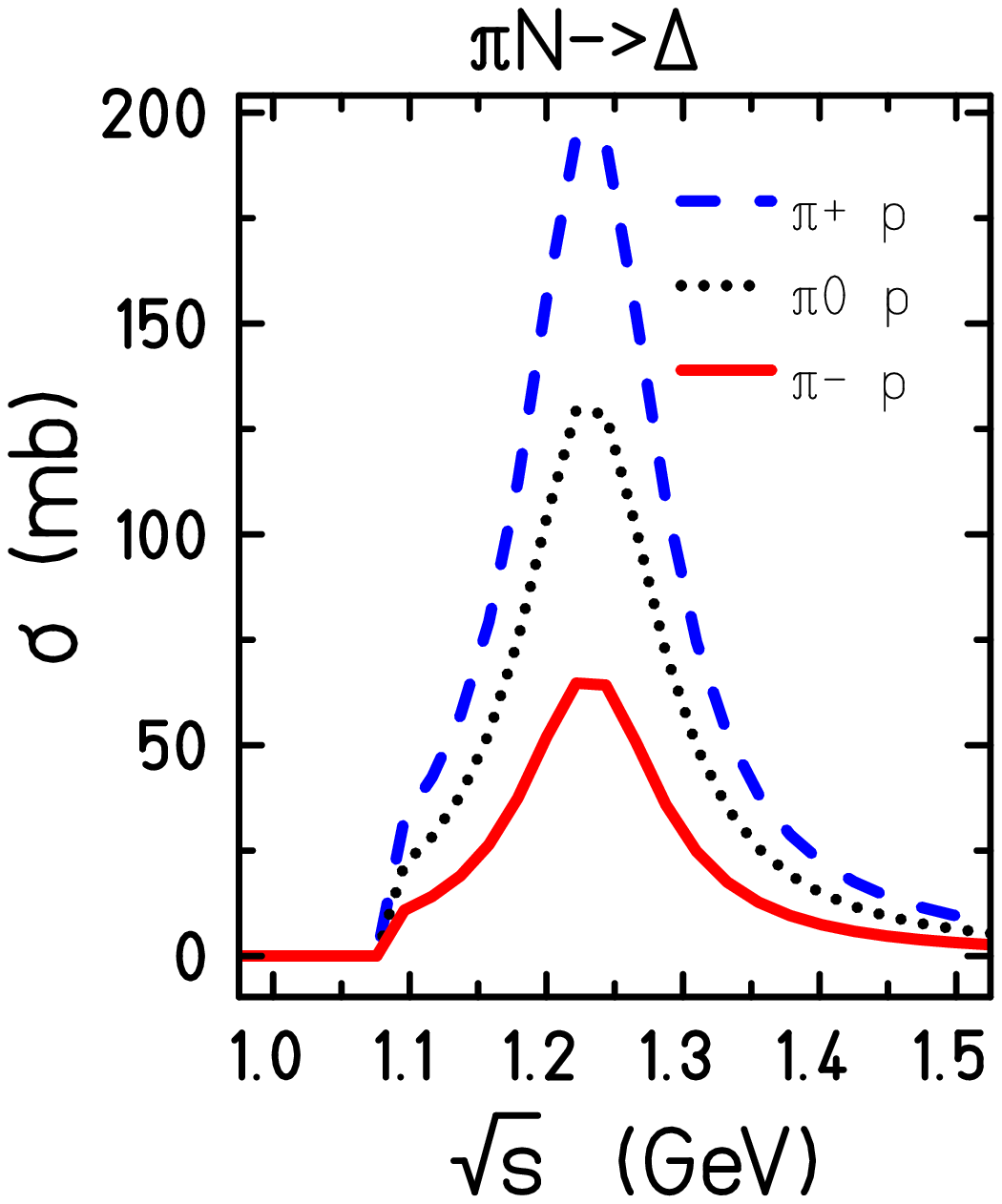,width=0.35\textwidth,angle=-0}}
\caption{Energy dependence of different cross sections used in
IQMD. } \vspace*{\figcapskip} \label{iqmd-xsections}
\end{figure}

We use the Isospin Quantum Molecular Dynamics model (IQMD)
\cite{iqmd,bass} in order to study the sensitivity of those ratios
on the variation of several cross sections. If the system is near
a thermal equilibrium, the influence of the cross sections is
negligible. The dominant cross section in the energy range under
study is besides the elastic nucleon-nucleon cross section the
inelastic reaction of two nucleons forming a $\Delta$ and the
corresponding inverse reaction $NN \leftrightarrow N\Delta$. The
produced $\Delta$ may also decay into a nucleon-pion pair and the
produced pion can again be absorbed by a nucleon forming a
$\Delta$: $\Delta \leftrightarrow N \pi$.

Figure~\ref{iqmd-xsections} presents on the left side the
parameterizations of the total pp (dashed) and pn cross sections
(dotted line) and their respective inelastic cross sections
forming a $\Delta$ (pp full line, pn thick dotted line). The
differences between the total and the inelastic cross section is
given by the elastic cross section. It should be noted that the
cross sections of the absorption of a $\Delta$ are related to
their production cross sections via detailed balance. The right
hand side shows the cross sections for the production of a
$\Delta$ by absorption of a pion. The inverse reaction, the
production of a pion by decay of a $\Delta$ is governed by its
decay width $\Gamma_\Delta$ (around 120 MeV in the pole). In the
following we use these parameterizations fitted to experimental
data of free nucleon-nucleon and pion-nucleon reactions and scale
them by a global factor. This has been done either by scaling all
cross sections or only by scaling a dedicated cross section. It
should be noted that the scaling of the cross section for $NN\to
N\Delta$ automatically scales the inverse reaction with the same
factor in order to assure detailed balance.

\begin{figure}[hbt]
\centerline{\psfig{file=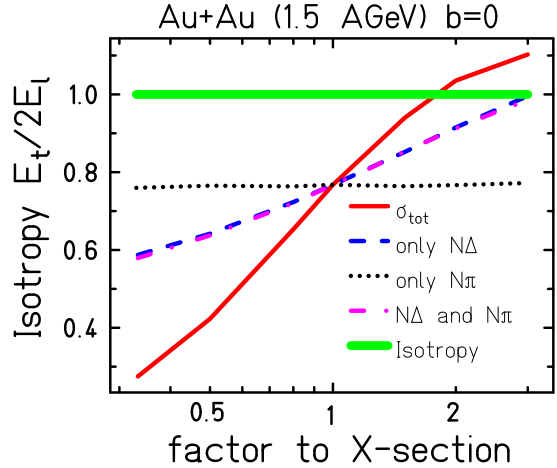 ,width=0.6\textwidth,angle=-0}}
\caption{Effect of the scaling of the cross sections on the isotropy ratio
of the momentum space distribution in central Au+Au collisions. }
\vspace*{\figcapskip}
\label{iqmd-isotropy}
\end{figure}

\section{Isotropy and pion production}
The first question is whether a dynamical equilibration in nuclear
phase space has been reached or not. One criterion is an isotropy
of the momentum distribution in the center of mass. We present in
Fig.~\ref{iqmd-isotropy} the ratio of transverse energy divided by
twice the longitudinal energy, $E_t/(2E_l)$, as a function a
multiplicative scaling factor to the nucleon cross sections. We
have chosen central collisions of a heavy system (Au+Au, $b$=0 fm
at 1.5 $A$GeV) which has a high degree of stopping. Smaller
systems do reach only lower values of this ratio. With the
standard cross section we do not reach unity (green thick line),
and a much higher cross section is needed to approach this value.
We find  a significant influence of the scaling of the inelastic
NN cross section (dashed line) which is related to the loss of
kinetic energy when producing a $\Delta$. An important part is due
to the elastic cross section, while the pion-induced cross section
(dashed line) does not show any sensitivity. No saturation is seen
when further increasing the cross sections which would also be a
possible signature of an equilibrium. We can thus conclude that a
global dynamical equilibrium is not reached.

\begin{figure}[hbt]
\centerline{\psfig{file=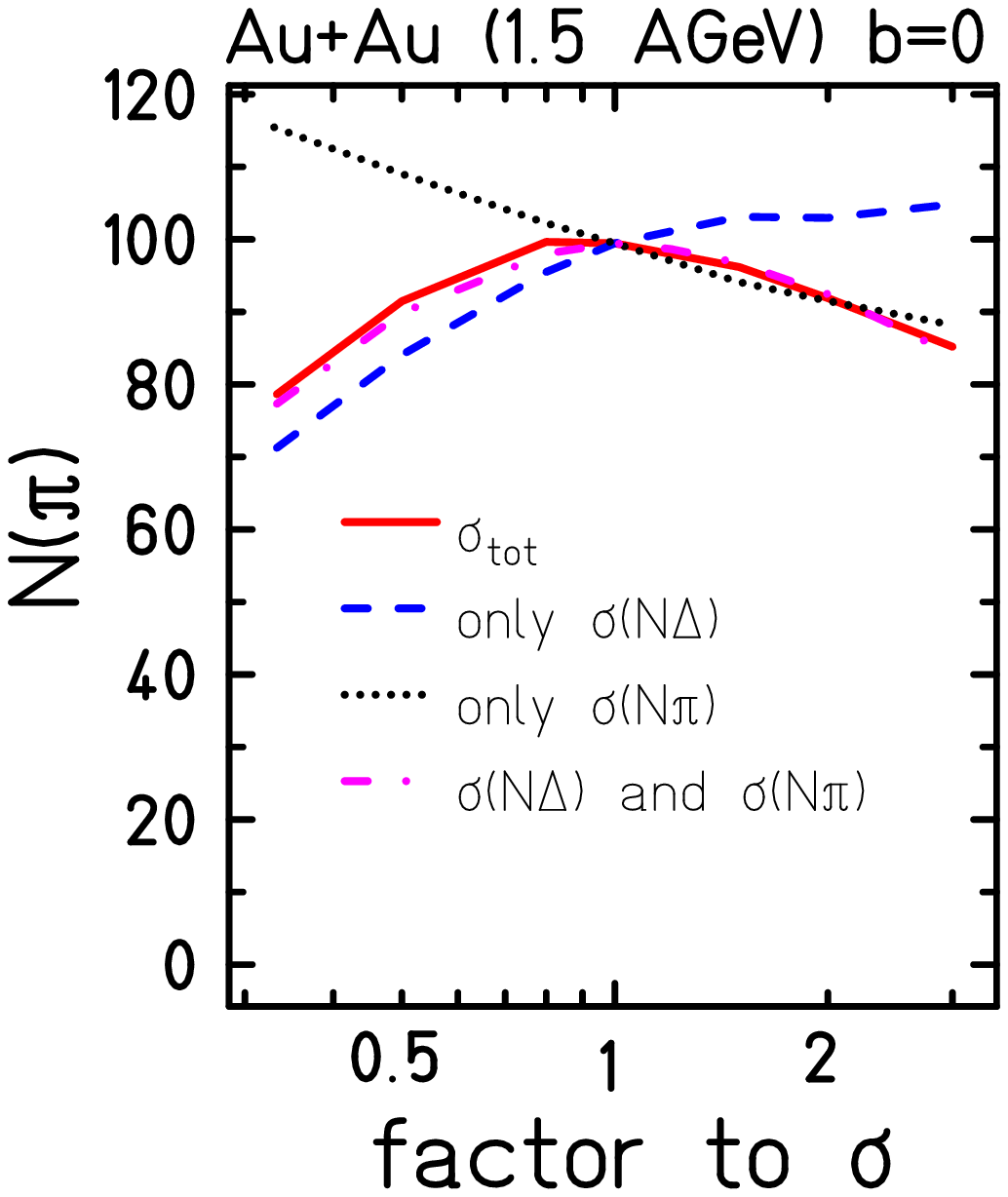
,width=0.4\textwidth,angle=-0}
\psfig{file=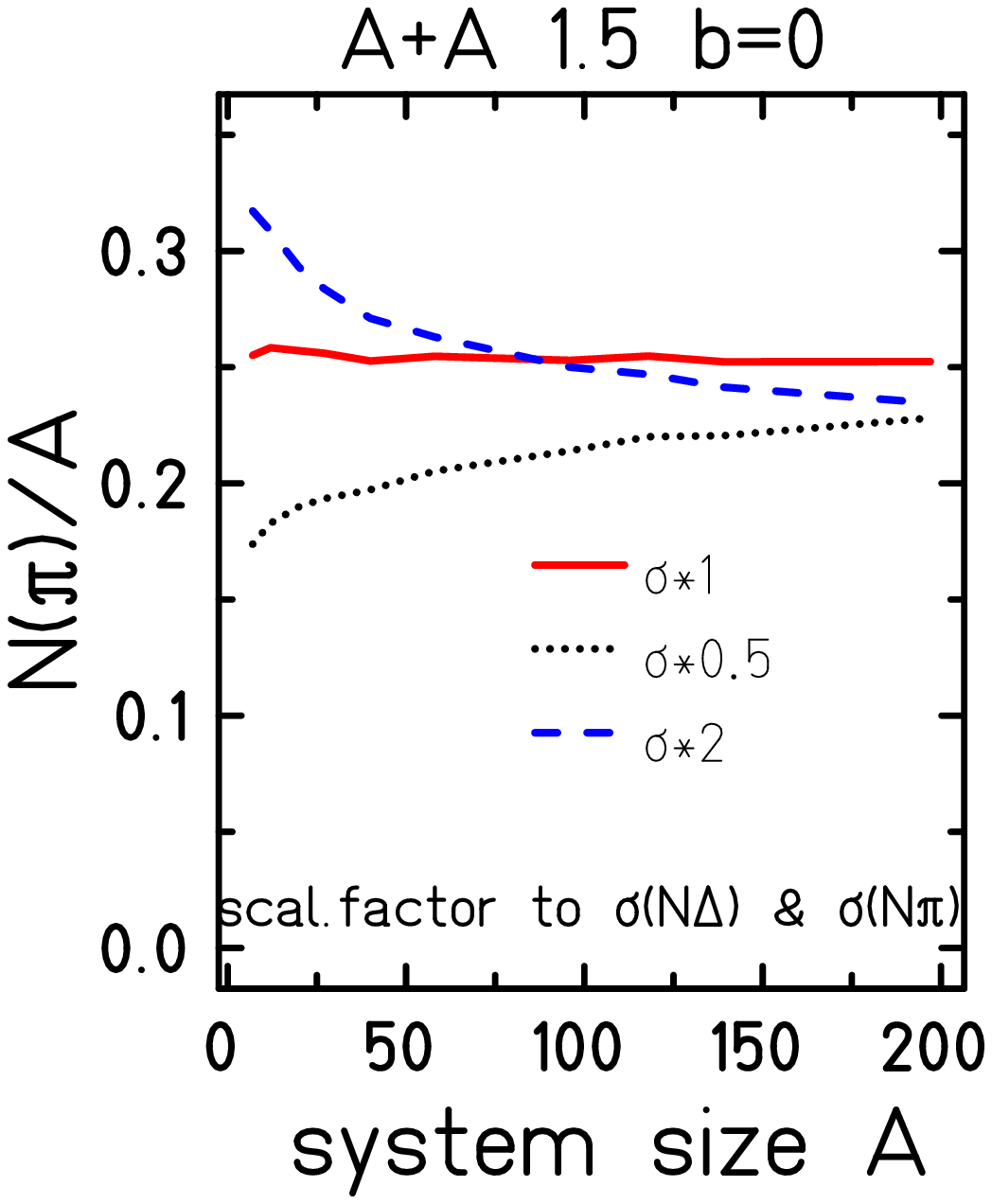 ,width=0.4\textwidth,angle=-0}
} \caption{Left:The total pion yield in
central Au+Au collisions at 1.5 $A$GeV as a function of the
scaling parameter to the cross sections.
Right:The total pion yield as function of the system size A
in central A+A collisions at 1.5 AGeV.
Different scaling parameters to the inelastic cross sections
have been employed } \vspace*{\figcapskip}
\label{iqmd-au-npi}
\end{figure}

Next we study the influence of the cross sections on the pion
yield, depicted on the left side of Fig.~\ref{iqmd-au-npi}, and no
saturation can be seen: The increase of the inelastic NN cross
sections ($NN \rightarrow N\Delta$ and its inverse reaction
$N\Delta\rightarrow NN $, dashed line) does not enhance the pion
yield significantly. Here we seem to reach a balancing of both
channels. On the other hand the increase of the $N \pi \rightarrow
\Delta$ cross section (dotted line, its reverse reaction is
governed by the  $\Delta$ decay width $\Gamma_\Delta$) decreases
the pion yield. This reaction reinserts the $\Delta$ and enhances
the probability of $N\Delta\rightarrow NN $, while the decay
$\Delta\to N\pi$ diminishes it. As a results, we change the
balance point of the $NN \leftrightarrow N\Delta$ system. When
applying the scaling factor to all cross sections (full line), we
obtain a maximum of the pion yield for a scaling factor of 1,
i.e.~for the present standard parametrization.

Let us now study the influence of this scaling parameter to the
system-size dependence as shown on the right side of
Fig.~\ref{iqmd-au-npi}. The pion yield has been divided by the
participant number, which is equivalent to the system size for
$b=0$. Only the standard parametrization (factor 1, full line)
assures a scaling of the pion number with the system size as
confirmed by experiment (see e.g.~\cite{reisdorf}), while other
scaling factors results in increasing or decreasing curves. This
special situation comforts the application of thermal models at
these low energies, since the pion yield can now be described as
an extensive variable.


\begin{figure}[hbt]
\centerline{\psfig{file=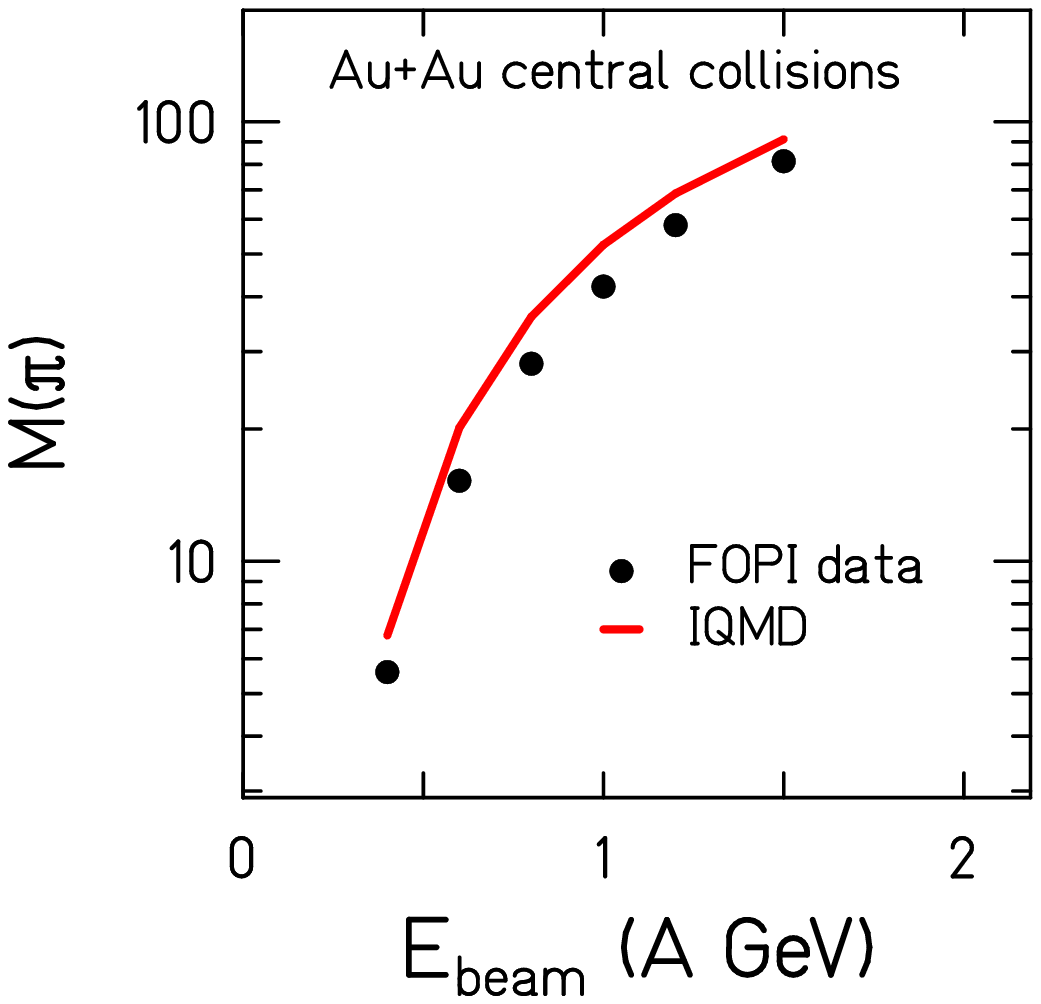,width=0.49\textwidth,angle=-0}
\psfig{file=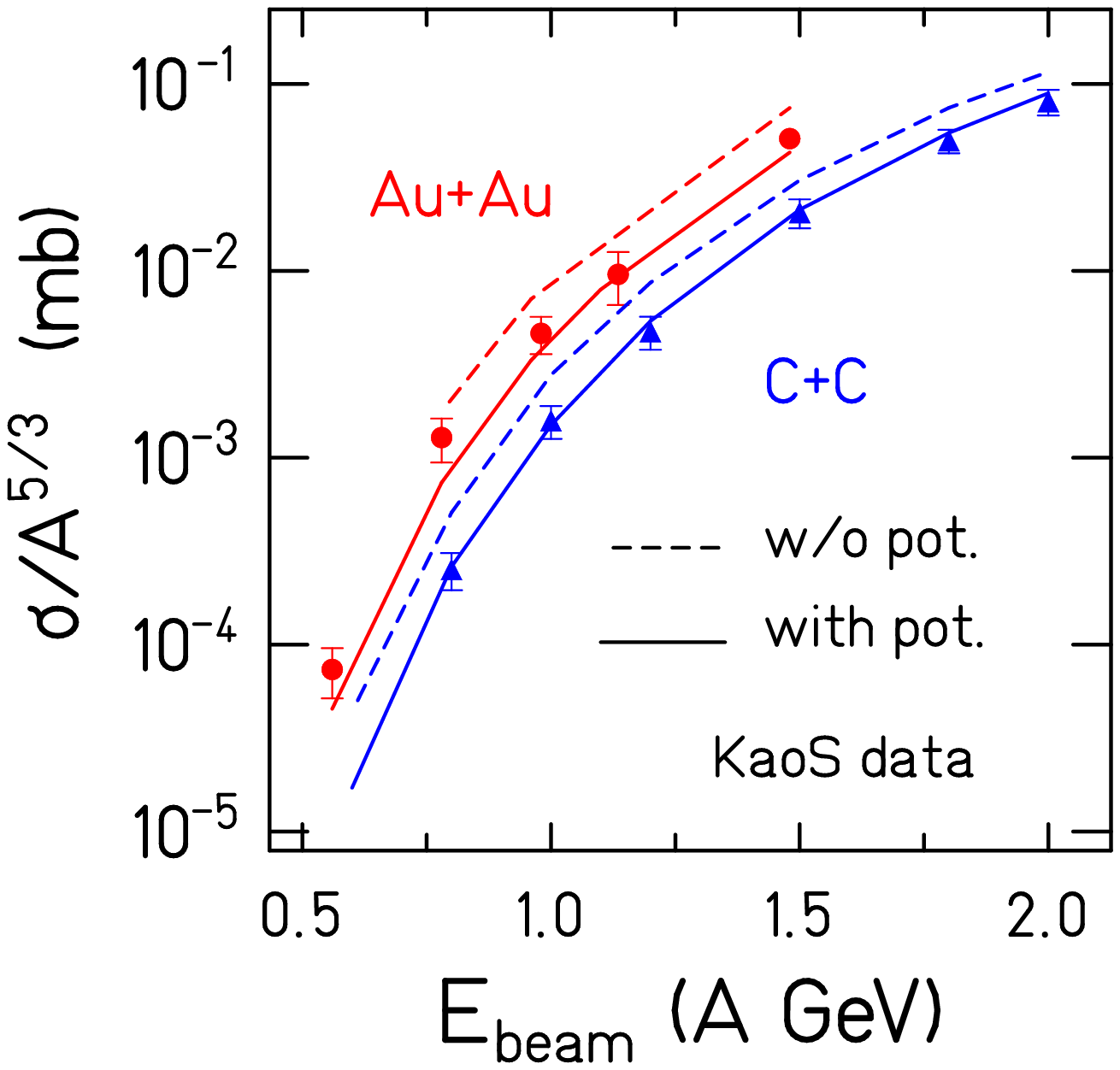,width=0.49\textwidth,angle=-0}
\hspace*{-0.08\textwidth} \vspace*{\figcapskip}}
\hspace*{-0.08\textwidth} \vspace*{\figcapskip}\\ 
\centerline{\psfig{file=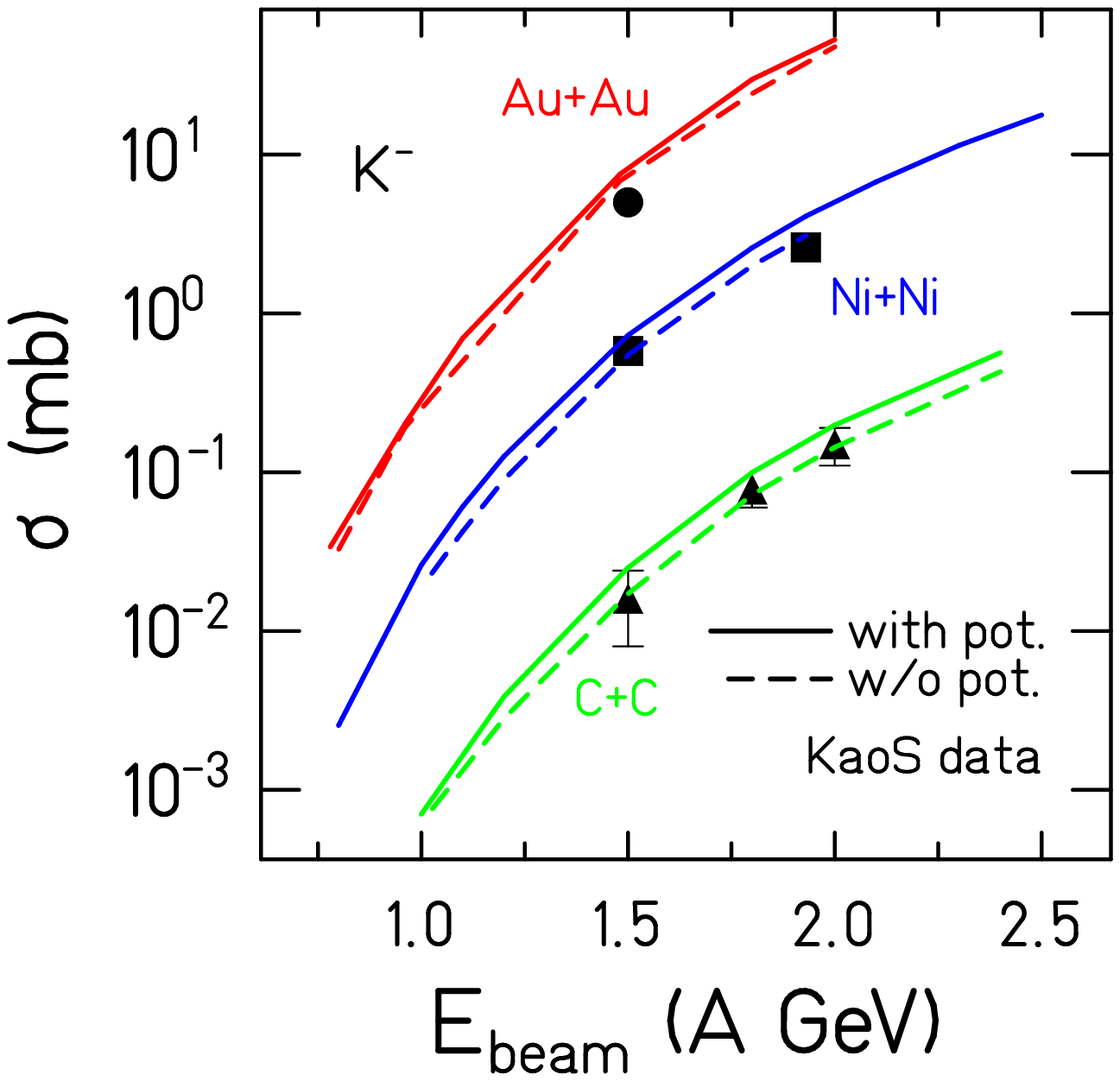,width=0.49\textwidth,angle=-0}
\vspace*{\figcapskip}}
\caption{Comparison of the excitation function of the pion yield
in central Au+Au collisions with FOPI data (top left) and of the $K^+$
(top right) and $K^-$ (bottom) yield in with experimental inclusive
data of the KaoS collaboration. } \vspace*{\figcapskip}
\label{iqmd-kaos-k}
\end{figure}

\section{Production of $K^+$ and $K^-$}

Before continuing we would like to state that IQMD calculations
are able to explain the experimental pion and kaon excitation
functions within about 10-20\% precision, as it is demonstrated in
Fig.~\ref{iqmd-kaos-k}. The experimental data stems from the FOPI
Collaboration for the pion \cite{reisdorf} and for the $K^+$ and
$K^-$ from KaoS Collaboration~\cite{FoersterPRC}.

The kaon production at this energy range is dominated by two-step
processes like $N_1N_2 \to N \Delta \quad N_3 \Delta \to NYK$
\cite{kaon1994,bigpr}. This chain of reactions is sensitive to the
yield of $\Delta$s and thus to $\sigma(NN \leftrightarrow
N\Delta)$. Since the production occurs in two steps, the
production probability depends also on the density reached in the
system. Therefore, the kaon yield is related to the degree of
stopping indicated in Fig.~\ref{iqmd-isotropy}. Furthermore, the
kaon yield depends on the life time of a $\Delta$, i.e.~the
duration when it is available for a collision.

\begin{figure}[hbt]
\centerline{\psfig{file=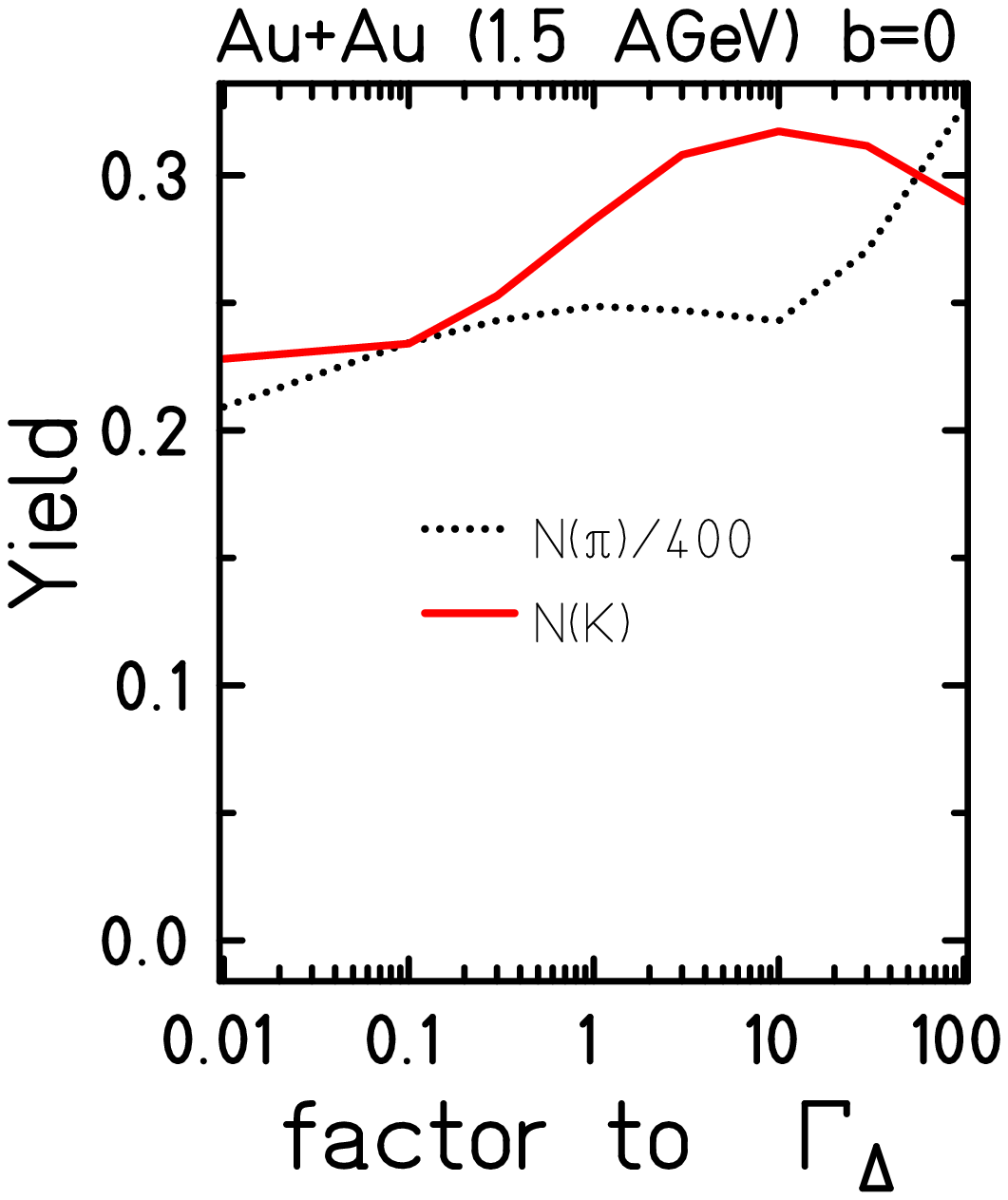,width=0.38\textwidth,angle=-0}
\psfig{file=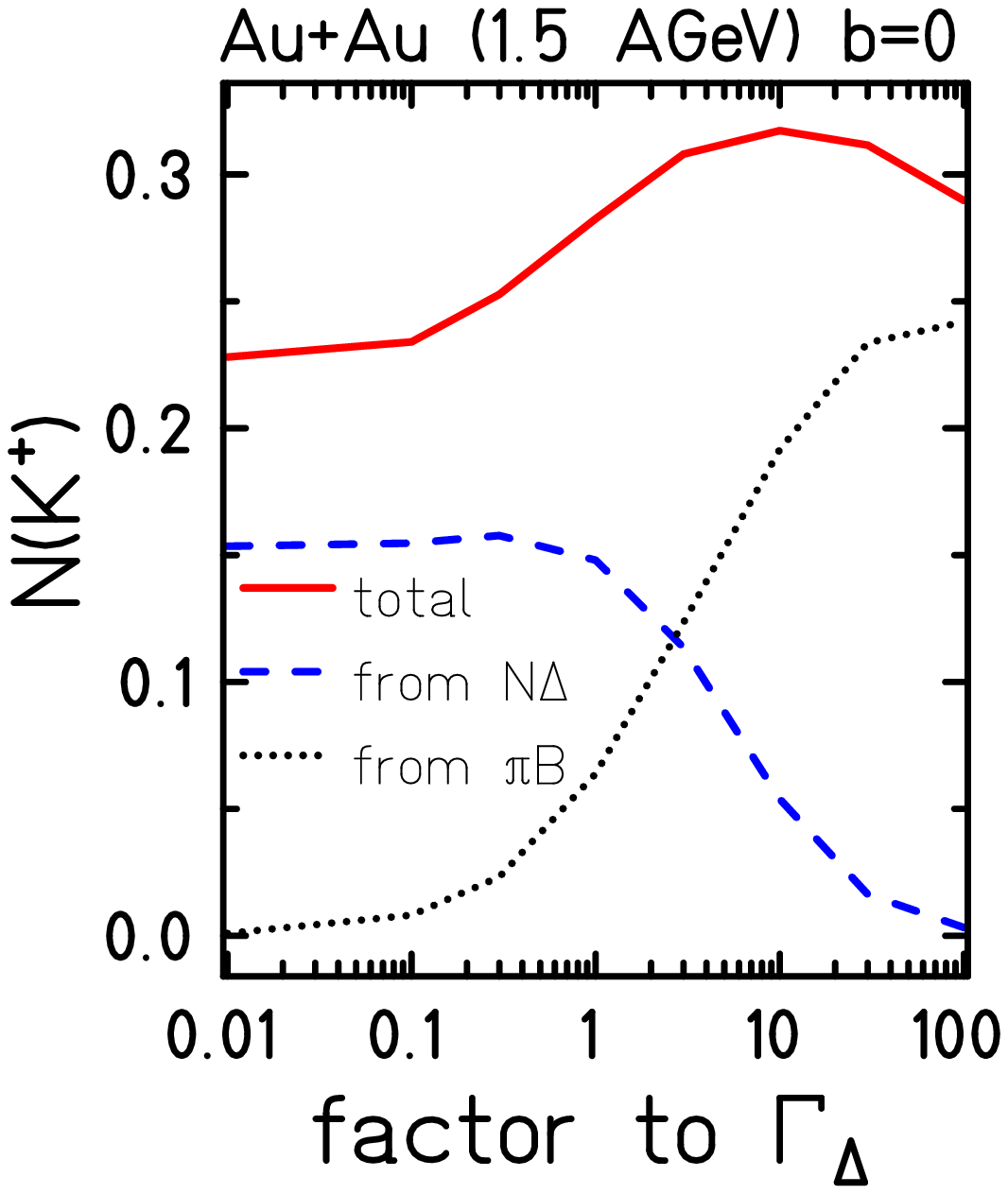,width=0.38\textwidth,angle=-0}}
\caption{Influence of a scaling factor to the $\Delta$-decay width
$\Gamma_\Delta$ on the final pion and kaon yield (left) and on the
contributions to the kaon yield (right)
} \vspace*{\figcapskip} \label{iqmd-gamma}
\end{figure}
The influence of the life time of the $\Delta$ on the kaon yield
is demonstrated in Fig.~\ref{iqmd-gamma}, left, applying an
multiplicative factor the decay width $\Gamma_\Delta$ which is
about 120 MeV at the pole. A factor less than 1 decreases the
width and enhances the life time. A larger factor reduces the life
time and thus penalizes the reaction $N\Delta\to NN$. the
influence of the life time is partially balanced by reabsorption
of the produced pion via the $N\pi\to\Delta$ reaction as can be
seen on the left side of Fig.~\ref{iqmd-gamma} where we show the
final yield of pions (dotted line, divide by a factor 400 to
compare with the kaons). Only for drastic changes of the decay
width the yield changes significantly. The yield of the kaons
(full line) changes quite moderately. The reason can be seen on
the right side of Fig.~\ref{iqmd-gamma}. The dominant channel
$N_1N_2 \to N \Delta$  and $ \quad N_3 \Delta \to NYK$ (dashed
line) is sensitive to the $\Delta$ lifetime and drops for larger
values of $\Gamma_\Delta$. However, this effect is to a large
extend balanced by the appearance of another production chain
$N_1N_2 \to N \Delta ,    \quad \Delta \to N\pi,  \quad N_3\pi\to
YK$ (dotted line). Only a small sensitivity of the total yield
(full line) remains.

\begin{figure}[hbt]
\centerline{\psfig{file=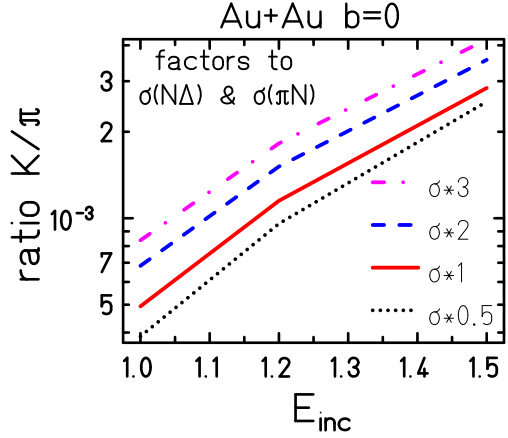
,width=0.55\textwidth,angle=-0}} \caption{Excitation function of
the $K/\pi$ ratio in Au+Au collisions for different scaling
factors applied to $NN\leftrightarrow N\Delta$ and
$\pi N\to \Delta$ at the same time.} \vspace*{\figcapskip} \label{iqmd-kpi}
\end{figure}
The fact that the kaon production depends strongly on a scaling of
the cross sections while the pions are only effected moderately
explains the strong influence of a scaling to the ratio of
$K^+/\pi$ shown in Fig.~\ref{iqmd-kpi}. The rising trend of the
curves is similar, only the absolut values vary with the scaling
factor. We would like to stress that the cross sections for the
dominant chain including a $N \Delta \to NYK$ are experimentally
unknown. Therefore, the absolute yield of the kaons contains a
large uncertainty.
Therefore, any conclusion from the absolute values of the
$K^+/\pi$ ratios in this energy range is quite difficult to
establish.

\begin{figure}[hbt]
\centerline{\psfig{file=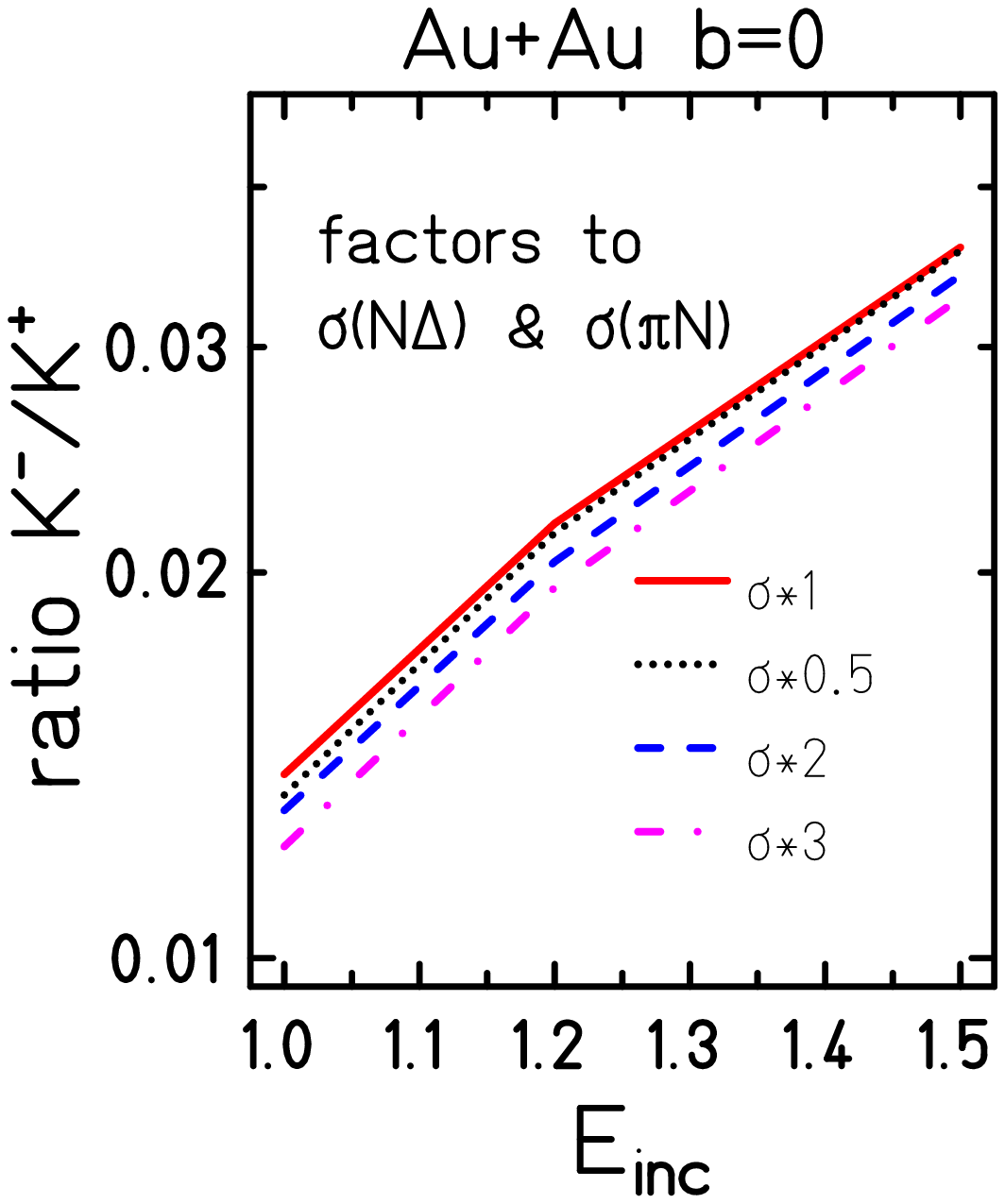 ,width=0.36\textwidth,angle=-0}
\psfig{file=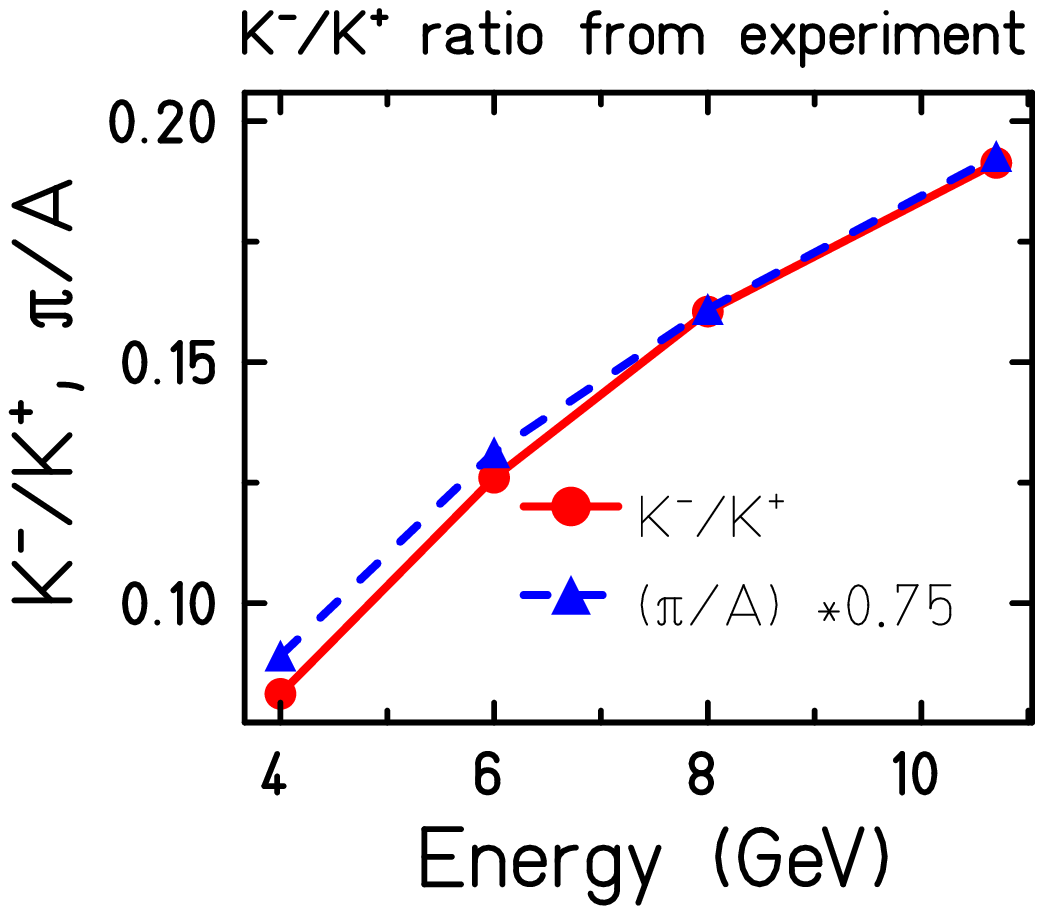 ,width=0.48\textwidth,angle=-0}}
\caption{Excitation function of the $K^-/K^+$ ratio in Au+Au collisions
for different scaling factors to the cross sections and comparison of
the $K^-/K^+$ ratio with the $\pi/A$ ratio. }
\vspace*{\figcapskip}
\label{iqmd-kmkp}
\end{figure}

The production of antikaons is dominated by charge exchange of the
hyperons: $\pi Y \leftrightarrow N \bar{K}$. Since the yield of
the hyperons is related to the yield of kaons via associate
production $BB\to NYK$ (which has a much lower threshold than the
direct production $BB\to NNK\bar{K}$), there exist a direct link
between the yield of kaons and that of antikaons via the hyperons.
Here we find first indications for the validity of a law of mass
action, as it has been reported in \cite{ho_s2000,law-of-mass}.
This idea was supported by IQMD calculations \cite{kminus} which
showed, that the yield of antikaons is rather insensitive on
scaling factors to the cross section of the reaction $\pi Y
\leftrightarrow N \bar{K}$ if we apply it to both directions. This
observation is interpreted as a dynamical balance between pions
and hyperons on one side and nucleons and antikaons on the other
side.

A scaling of other cross sections (like $N\Delta$ or $N\pi$) does
not effect that link between kaons and antikaons. This is
demonstrated on the left side of Fig.~\ref{iqmd-kmkp}, where a
scaling of the cross sections only causes a small effect on the
ratio of $K^-/K^+$. The other parameter governing this ratio is
the pion yield, as shown on the right side of
Fig.~\ref{iqmd-kmkp}: experimental data of the ratio $K^-/K^+$
show the same excitation function as the experimental pion yield,
scaled with a factor of the participant size \cite{ho_s2000}.

\begin{figure}[hbt]
\centerline{\psfig{file=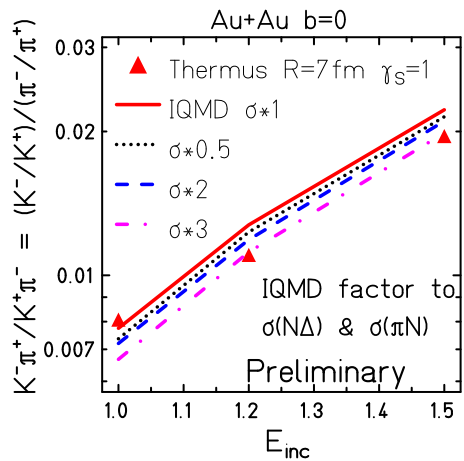
,width=0.5\textwidth,angle=-0}} \caption{Excitation function of
the $(K^-/K^+)/(\pi^-/\pi^+)$ double ratio in Au+Au collisions for
different scaling factors to the cross sections and in comparison
to THERMUS results\cite{Spencer}} \vspace*{\figcapskip}
\label{iqmd-thermus}
\end{figure}

The values of this ratio correspond to the values obtained by
thermal models like THERMUS \cite{thermus}.
Figure~\ref{iqmd-thermus} shows the comparison of the ratio
$(K^-/K^+)/(\pi^-/\pi^+)$ between THERMUS (triangles, compiled
from $K^-/\pi^-$ and $K^+/\pi^+$ taken from \cite{Spencer}) with
IQMD calculations (lines). Since the ratio $\pi^-/\pi^+$ is mostly
sensitive to the isospin composition of projectile and target, the
excitation function of the ratio $(K^-/K^+)/(\pi^-/\pi^+)$
corresponds mainly to the ratio $K^-/K^+$. This ratio turn out to
be independent from a scaling of the cross sections and the values
are comparable to the values obtained with THERMUS. This
observable demonstrates well the validity of the law of mass
action. Due to this relation, a description with thermal models
appears applicable.

\section{Conclusion}
We have used the IQMD model to study the influence of a scaling of
cross sections on the yield of pions, kaons and antikaons. The
pion yield is slightly effected by a scaling of the cross sections
and yields a maximum for the standard parametrization obtained
from the free experimental cross sections. The scaling of the pion
yield with the participant number can only be assured when no
scaling of the cross sections is done. The ratio of $K/\pi$
depends strongly on the scaling of the cross sections but its
absolute value suffer from the uncertainty in unknown production
channels of the kaon. The ratio $K^-/K^+$ reflects nicely the
properties of the law of mass action. Consequently, thermal models
are applicable for some but not all ratios is this energy range.


\end{document}